\documentclass[12pt]{article}
\usepackage{amsmath, latexsym, epsfig, graphicx, rotating, fancyhdr, tabularx, afterpage}
\usepackage{axodraw,epsfig,color}
\usepackage{xspace}
\newcommand{\sss}{\scriptscriptstyle}
\newcommand{\mw}{m_{\sss W}} 
\newcommand{\mz}{m_{\sss Z}} 

\newcommand{\nll}{\nonumber \\}
\newcommand{\be}{\begin{equation}}
\newcommand{\ee}{\end{equation}}
\newcommand{\bqa}{\begin{eqnarray}}
\newcommand{\eqa}{\end{eqnarray}}

\newcommand{\sw}{s^2_\smallw}
\newcommand{\cw}{c^2_\smallw}
\newcommand{\stw}{s_\smallw}
\newcommand{\ctw}{c_\smallw}
\newcommand{\swb}{\bar{s}^2_\smallw}
\newcommand{\cwb}{\bar{c}^2_\smallw}
\newcommand{\xt}{x_t}
\newcommand{\mt}{m_t}
\newcommand{\smallh}{{\scriptscriptstyle H}}
\newcommand{\smallw}{{\scriptscriptstyle W}}
\newcommand{\mh}{m_\smallh}
\newcommand{\rhoone}{\rho^{(1)}}
\newcommand{\rhotwo}{\rho^{(2)}}

\providecommand{\MCSANC}{{\textsc{MCSANC}}\xspace}

\begin{document}

\begin{center}

{\LARGE {Update of the \MCSANC Monte Carlo Integrator, v.1.20}} 

\vspace*{1cm}
A.\,Arbuzov$^{b}$, D.\,Bardin$^{a}$\/\thanks{e-mail: sanc@jinr.ru}, S.\,Bondarenko$^{b}$, 
P.\,Christova$^{a,}$, L.\,Kalinovskaya$^{a}$, U.\,Klein$^{d}$, V.\,Kolesnikov$^{a}$, 
L. Rumyantsev$^{a,c}$, R.,\,Sadykov$^{a}$, A.\,Sapronov$^{a}$.

$^{a}$ Dzhelepov Laboratory of Nuclear Problems, JINR,      
        Joliot-Curie str. 6, 141980 Dubna, Russia;         

$^{b}$ Bogoliubov Laboratory of  Theoretical Physics, JINR,  
        Joliot-Curie str. 6, 141980 Dubna, Russia;         

$^{c}$ Institute of Physics, Southern Federal University, Rostov-on-Don,
 344090 Russia;

$^{d}$ University of Liverpool, Liverpool, UK.
\end{center}
\abstract{
This article  presents new features of the \MCSANC~{\tt v.1.20} program, a Monte
Carlo tool for calculation of the next-to-leading order electroweak and QCD
corrections to various Standard Model processes.  
The extensions concern implementation of  Drell--Yan-like
processes and include a systematic treatment of the photon-induced contribution in
proton--proton collisions and electroweak corrections beyond NLO approximation.
There are also technical improvements such as calculation of the
forward-backward asymmetry for the neutral current Drell--Yan process. The updated
code is suitable for studies of the effects due to EW and QCD
radiative corrections to Drell--Yan (and several other) processes at the LHC
and for forthcoming high energy proton--proton colliders. 
}

\tableofcontents

\section{Introduction}
\label{Introd}

The forthcoming LHC data on Drell--Yan (DY) processes 
allows to access final states with very high invariant di-lepton masses, 
where the photon-induced contributions 
become substantial relative to the standard quark--antiquark annihilation
sub-processes. An accurate estimate of these contributions for
hypothetical high mass resonance searches requires the inclusion into the theory predictions
sub-processes with photons in the initial $pp$ state such as $q \gamma \to q' \ell^{\pm} \nu_\ell$,
$q \gamma \to q \ell^+ \ell^-$ and {$\gamma \gamma \to \ell^+ \ell^- $. 

Corrections to the neutral current Drell--Yan (NC DY) cross-section due to
photon-induced process $ \gamma \gamma \to \ell^+ \ell^- $ can reach up to
$10-20 \% $ for high invariant mass $ M_{\ell^+\ell^-} $ with a choice of
kinematic cuts typical for LHC experimental analysis.
A first evidence of this kind of background was found  by the ATLAS
Collaboration in high mass NC DY cross section measurements \cite{Aad:2013iua}.

Photon-induced Drell--Yan processes were carefully investigated in many papers, see for example 
\cite{Brensing:2007qm} and  \cite{Arbuzov:2007kp}.

In the TeV region of invariant masses the higher order two-loop electroweak
(EW) and QCD leading terms are important for estimation of the theoretical
uncertainties in different EW schemes.

With this paper we continue the series of works
\cite{Bardin:2012jk},~\cite{Bondarenko:2013nu}
dedicated to the development of \MCSANC, a Monte Carlo tool based on the {\sc SANC}
modules~\cite{Andonov:2008ga}\footnote{This paper uses the same nomenclature
as introduced in \cite{Arbuzov:2007kp}}.
We present an update of the integrator up to {\tt v.1.20}
with inclusion of the aforesaid corrections relevant for DY processes at the 
LHC at $\sqrt{s} = 13$\,TeV.
We briefly review the implementation into the framework of the \MCSANC~{\tt v.1.20} tool
the following three new options:
\begin{itemize}
\item photon-induced contributions. The implemented processes are:
\begin{itemize}
 \item{$q \gamma \to q' \ell^{\pm} \nu_\ell$} (for CC DY), 
 \item{$q \gamma \to q \ell^+ \ell^-$} (for NC DY), 
 \item{$\gamma \gamma \to \ell^+ \ell^- $} (for NC DY);
\end{itemize}

\item Leading in $G_\mu\mt^2 $ two-loop EW and mixed EW$\otimes$QCD radiative corrections; 

\item forward-backward asymmetry $A_{FB}^{ff}$.
\end{itemize}

This paper is organized as follows.
In Section \ref{PIP} we describe the implementation of the photon-induced processes.
Section \ref{TLEC} is devoted to the accounting of
higher order radiative corrections using the $\rho$ parameter at two loops.
Results of our estimates of photon-induced processes,
higher order radiative corrections, and forward-backward asymmetry $A_{FB}^{ff}$,
as well as comparison between \MCSANC~ {\tt v.1.20} and~\cite{Brensing:2007qm} 
and ~\cite{Dittmaier:2009cr}
are presented in Section {\ref{NR}}.
A brief conclusion is given in Sect.~\ref{conclusion}.

\section{Photon-induced processes}
\label{PIP}

The introduction of photon-induced corrections into the {\sc SANC} environment is presented
in paper \cite{Arbuzov:2007kp} where the ${\overline{\mathrm{MS}}}$ subtraction scheme is realized.

The DIS subtraction scheme can be realized using the sum of the following  subtraction terms:
\begin{equation}
  \left(\delta_{1}^{\overline{\mathrm{MS}}}+\delta_{1}^{\text{DIS}}\right) 
 +\left(\delta_{2}^{\overline{\mathrm{MS}}}+\delta_{2}^{\text{DIS}} \right).
\label{dlts}
\end{equation}
Here $\delta_{1}^{\overline{\mathrm{MS}}}=\delta_{1}(c_1),\quad \delta_{2}^{\overline{\mathrm{MS}}}=\delta_{2}(c_1)$ 
and the corresponding structure function are given 
by Eqs.~(6) and (8) in paper~\cite{Arbuzov:2007kp}. The structure function
\begin{equation}
D_{q^{'}_i\gamma}^{\text{DIS}}(x) = - \log \frac{Q^2 (1-x)}{x m_q^2} (x^2+(1-x)^2) - 1 + 8 x - 8 x^2
\end{equation}
should be used within the kinematics~(5) of the above paper. 
Note that expression~(\ref{dlts}) is for NC DY processes 
while only the first two terms in parentheses are present for CC DY ones.

The $\delta_{2}^{\text{DIS}}$ contribution is calculated within the following kinematics
\begin{eqnarray}
\begin{split}
\delta_{2}^{\text{DIS}} &=
\sum_{q_i} \int_{2m_f}^{\sqrt{s_0}} dM \int_{-\infty}^{+\infty} dY \int_0^1 dx_3 \int_{-1}^1 d\cos{\hat{\vartheta}} \;\;
q_i\left(x_1,\mu_F^2\right)\gamma\left(x_2,\mu_F^2\right) \\
\label{threeprime}
&\times \frac{d\hat{\sigma}^{\gamma\gamma}(M,\cos{\hat{\vartheta}})}{d\cos{\hat{\vartheta}}}
\frac{2M}{x_3s_0}\Theta\left(1-x_1\right)
\frac{\alpha}{2\pi}Q_{q_i}^2
\left( C_{\gamma q_i}^{\text{DIS}}(x_3) - \delta(1-x_3)\int_0^1 C_{\gamma q_i}^{\text{DIS}}(z) dz \right)\\
&=\sum_{q_i} \int_{2m_f}^{\sqrt{s_0}} dM \int_{-\infty}^{+\infty} dY \int_0^1 dx_3 \int_{-1}^1 d\cos{\hat{\vartheta}}
\;\;
\frac{2M}{s_0}\\
&\times
\left(\frac{q_i\left(x_1,\mu_F^2\right)}{x_3}\Theta\left(1-x_1\right) 
- q_i\left(\bar{x}_1,\mu_F^2\right)\Theta\left(1-\bar{x}_1\right) \right)\\
&\times \gamma\left(x_2,\mu_F^2\right) \frac{d\hat{\sigma}^{\gamma\gamma}
(M,\cos{\hat{\vartheta}})}{d\cos{\hat{\vartheta}}} D_{\gamma q_i}^{\text{DIS}}(x_3),
\end{split}
\end{eqnarray}
where the coefficient function reads
\begin{equation}
C_{\gamma q_i}^{\text{DIS}}(x) = -
\left(
\frac{1+x^2}{1-x}\left(\log{\frac{1-x}{x}}-\frac{3}{4}\right)
+\frac{9+5x}{4}
\right),
\nonumber
\end{equation}
and
\begin{equation}
\begin{split}
&M = \sqrt{x_1x_2x_3s_0}, \qquad Y = \frac{1}{2}\log{\frac{x_1x_3}{x_2}},\\
&x_1 =\frac{M}{\sqrt{s_0}}\frac{e^{Y}}{x_3},\quad
\bar{x}_1 = x_1x_3, \quad x_2 = \frac{M}{\sqrt{s_0}}e^{-Y},
\end{split}
\nonumber
\end{equation}
where $\sqrt{s_0}$ is the c.m.s. energy of colliding protons. Regularization of function 
$C_x$ at $x \rightarrow 1$ is performed by plus-prescription explicitly shown in (\ref{threeprime}).

Another sub-process with photons in the initial state is $\gamma\gamma \to \ell^+ \ell^-$ which
contributes to NC DY only. Here we give only the cross section of the process
in the massless case,
\bqa
\hat{\sigma}^{\gamma\gamma,Born} = \frac{\alpha^2 \pi}{s}\left(\frac{t}{u}+\frac{u}{t}\right),
\eqa
and in the massive case,
\bqa
\hat{\sigma}^{\gamma\gamma,Born} = \hspace*{-5mm}
&& \frac{\alpha^2 \pi}{s}
 \Bigl\{
 \frac{1}{1-\beta_l\cos\vartheta}
    \left[1+\beta_l\cos\vartheta+\frac{4m_f^2}{s} 
 \left(1-\frac{1+\beta_l^2}{1-\beta_l\cos\vartheta}\right)    \right]
 \nll &&
 +\frac{1}{1+\beta_l\cos\vartheta}
    \left[1-\beta_l\cos\vartheta+\frac{4m_f^2}{s}
\left(1-\frac{1+\beta_l^2}{1+\beta_l\cos\vartheta}\right)\right]
 \Bigr\},
\eqa
where $\displaystyle \beta_l =\beta(s,m_l^2,m_l^2) = \sqrt{1-4\frac{m_l^2}{s}}$
and $\vartheta$ is the angle between the photon and outgoing lepton momenta
in the center-of-mass system.

\section{Leading two-loop electroweak corrections}
\label{TLEC}

In \MCSANC~ {\tt v.1.20} we follow the recipe introduced in 
Refs.~\cite{Fleischer:1993ub} and ~\cite{Fleischer:1994cb}, 
and later well described in Ref~\cite{Bardin:280836}.

The $\rho$ parameter is defined as the ratio
 of the neutral current to
charged current amplitudes at zero momentum transfer,
see for example ~\cite{Fleischer:1993ub}:
\begin{equation}
\rho=\frac{G_{NC}(0)}{G_{CC}(0)}=\frac{1}{1-\Delta\rho}\,,
\label{def-rho}
\end{equation} 
where $G_{CC}(0)=G_\mu$ is the Fermi constant defined from the $\mu$-decay width,
and the quantity $\Delta\rho$ is treated perturbatively
\begin{equation}
\Delta\rho=\Delta\rhoone+\Delta\rhotwo+\ldots
\end{equation}

Expanding (\ref{def-rho}) up to quadratic terms $\Delta\rho^2$, we have
\begin{equation}
\rho=1+\Delta\rho+\Delta\rho^2\,.
\label{rho-upto-quadro}
\end{equation}
The contribution to $\Delta\rho$, leading in $G_\mu\mt^2$ NLO EW,
 is explicitly given by
\begin{equation}
\Delta\rhoone\Big|^{G_\mu}=3\xt=\frac{3\sqrt{2}G_\mu\mt^2}{16\pi^2}\,.
\label{def-rho-one}
\end{equation}

At the two-loop level, the quantity $\Delta\rho$ contains two  contributions:
\begin{equation}
\Delta\rho=N_c\xt\,\left[1+\rhotwo\,\left(\mh^2/\mt^2\right)\,\xt\,\right]\,
            \left[1-\frac{2\alpha_s(\mz^2)}{9\pi}(\pi^2+3)\right].
\end{equation}
They consist of the following:

i) two-loop EW part at ${\cal O}(G_\mu^2)$, second term in the first square 
brackets~\cite{Barbieri:1992nz},~\cite{Fleischer:1993ub} and~\cite{Fleischer:1994cb}
with $\rhotwo$ given in Eq. (12) of~\cite{Fleischer:1994cb}
(actually, after discovery of the Higgs boson
and determination of its mass, it has become
sufficient to use the low Higgs mass asymptotic, Eq.~(15), of~\cite{Fleischer:1994cb}); 

ii) mixed two-loop EW$\otimes$QCD at ${\cal O}(G_\mu\alpha_s)$,
the second term in the second square 
brackets, see in ~\cite{Djouadi:1987gn}--\cite{Djouadi:1987di} for further details.

From Eq.~(\ref{def-rho}), using intermediate vector boson propagators 
$\sim 1/(Q^2+M^2_V)$, we derive:
\begin{equation}
\rho=\frac{\mw^2}{\cwb\mz^2}\,,
\label{def-cwb}
\end{equation}
where we have introduced a new parameter $\cwb$ to distinguish from the usual $\cw$ 
for which we maintain the meaning
$\cw=\mw^2/\mz^2$ to be valid to all perturbative orders. At the lowest order (LO)
\begin{equation}
\rho^{(0)}=\frac{\mw^2}{\cw\mz^2}=1\,.
\end{equation}
From Eq.(\ref{def-cwb}) we have:
\begin{equation}
\cwb=\frac{\mw^2}{\rho\mz^2}=(1-\Delta\rho)\,\cw\,.
\label{cwb}
\end{equation}

The universal higher order (h.o.) corrections, leading in $G_\mu\mt^2$,
may be taken into account via the following replacements:
\bqa
&&\alpha_{G_\mu}\to \alpha_{G_\mu} \frac{\swb}{\sw}\,,
\label{basic:replacement}
\\
&&\sw\to\swb\equiv\sw+\Delta\rho\,\cw\,,\quad\cw\to\cwb\equiv 1-\swb=(1-\Delta\rho)\,\cw\,
\label{eq:replacements}
\eqa
in the LO expression for NC DY cross section
(see discussion after Eq.(3.48)) of~\cite{Dittmaier:2009cr}).

As was argued in Refs. \cite{Consoli:1989fg},
\cite{Consoli:1989pc} and \cite{Fleischer:1993ub}, 
this approach correctly reproduces terms up to ${\cal O}(\Delta\rho^2)$.

Given the replacements in Eq.(\ref{eq:replacements}), we get the
following contributions of h.o. EW corrections
to the scalar form factors 
\footnote{For the definition of the scalar form factors see Eqs.(29)--(30) in \cite{Andonov:2004hi}.}
of the invariant amplitude. 
In the $G_\mu$-scheme, 
\bqa
\alpha_{G_\mu}=\frac{\sqrt{2}G_\mu\mw^2\sw}{\pi}\,,
\eqa
the form factor ${\cal{F}}_{\gamma}$ of $\gamma$ exchange effectively contains the factor $\sw$
\begin{equation}
\alpha_{G_\mu}{\cal{F}}_{\gamma} \to \alpha_{G\mu}{\swb}/\sw= 
\alpha_{G\mu}\left(1+\frac{\cw}{\sw}\Delta\rho\,\right),
\label{FFAexc}
\end{equation}
while four form factors of $Z$ exchange, 
${\cal{F}}^{LL}_{\sss Z},~{\cal{F}}^{LQ}_{\sss Z},~{\cal{F}}^{QL}_{\sss Z},~{\cal{F}}^{QQ}_{\sss Z}$, 
contain a common factor $1/\cw$. We consider ${\cal{F}}^{LL}_{\sss Z}$ as an example
\bqa
\alpha_{G_\mu}\kappa{\cal{F}}^{LL}_{\sss Z} = \alpha_{G_\mu}\frac{1}{4\sw\cw} {\cal{F}}^{LL}_{\sss Z}.
\eqa
Since the coupling factor $\alpha_{G_\mu}/\sw$ does not receive universal corrections, 
as follows from Eq.(\ref{FFAexc}), we should insert only factors which come from $1/\cw$,
\bqa
&&\alpha_{G_\mu}\frac{1}{4\sw\cw } {\cal{F}}^{LL}_{\sss Z}\to
\alpha_{G_\mu}\frac{1}{4\sw\cwb }=
\alpha_{G_\mu}\frac{1}{4\sw\cw }(1+\Delta\rho+\Delta\rho^2).
\label{FFZexc}
\eqa
In addition to Eq.(\ref{FFZexc}), form factors ${\cal F}^{LQ,QL}_{\sss Z}$ of $Z$ 
exchange contain the factor $\sw$
and form factor  ${\cal F}^{QQ}_{\sss Z} $ contains the factor $s_{\sss W}^4$.
Therefore, form factors at NNLO order read:
\bqa
 {\cal{F}}_{\gamma}^{QQ}  &=& 1 + \frac{\ctw^2}{\stw^2}\Delta\rho,         \nll
 {\cal{F}}_{\sss Z}^{LL}  &=& 1 + \Delta\rho+ \Delta\rho^2,                \nll
 {\cal{F}}_{\sss Z}^{LQ}  &=& (1 + \Delta\rho+ \Delta\rho^2)(1+\frac{\ctw^2}{\stw^2}\Delta\rho),\nll
 {\cal{F}}_{\sss Z}^{QL}  &=& (1 + \Delta\rho+ \Delta\rho^2)(1+\frac{\ctw^2}{\stw^2}\Delta\rho),\nll
 {\cal{F}}_{\sss Z}^{QQ}  &=& (1 + \Delta\rho+ \Delta\rho^2)(1+\frac{\ctw^2}{\stw^2}\Delta\rho)^2.
\label{ffll}
\eqa

To avoid double counting one should remove the leading NLO EW contribution (\ref{def-rho-one})
from the terms linear in $\Delta\rho$:
$\Delta\rho \longrightarrow \left( \Delta\rho -\Delta\rhoone\Big|^{G_\mu}\right)$.

Eqs.(\ref{ffll}) were realized in new {\tt modules} which compute the
differential cross section
contributions taking into account terms up to the order ${\cal O}(\Delta\rho)$ ({\tt iho=1})
or ${\cal O}(\Delta\rho^2)$ ({\tt iho=2}).

We have shown analytically that the results obtained in this way 
agree with the corresponding expressions derived in \cite{Dittmaier:2009cr}.

\section{Numerical results}
\label{NR}

\subsection{Photon-induced processes}
\label{NPIP}

In Tables ~\ref{TableDKl} and \ref{TableDKll} we present the LO inclusive 
cross sections for the process $pp \to e^+\nu_e X$
and the photon-induced contributions 
$\displaystyle\delta_{q/\bar q\gamma}=\displaystyle\frac{\sigma_{q/\bar q\gamma}}{\sigma_0}$
for different ranges of the lepton pair transverse mass $M_{T, l\nu}$ and of the
transverse momentum $p_{T, l}$, respectively. For the sake of comparison we used the setup 
and input parameters from the paper \cite{Brensing:2007qm}.
Results of \MCSANC (first line) are compared to the ones presented 
in Table~1 of~\cite{Brensing:2007qm} (second line).
The lowest order cross sections are given in picobarns, the correction factors are shown in \%.
The numbers illustrate good agreement within the statistical errors of Monte Carlo integration.

\begin{table}[!h]
\begin{center}
\begin{tabular}{|c|c|c|}
\hline
$M_{T, l\nu}$/GeV &   $\sigma_0$/pb &   $\delta_{q/\bar q\gamma}/\%$\\
\hline
 50-$\infty$    &   4495.8(1)     &    0.047(3) \\
                &   4495.7(2)     &    0.052(1) \\
\hline
100-$\infty$    &   27.590(1)     &    0.11(1) \\
                &   27.589(2)     &    0.12(1)  \\
\hline
 200-$\infty$   &   1.7907(1)     &    0.24(1) \\
                &   1.7906(1)     &    0.25(1)  \\
\hline
 500-$\infty$   &   0.084696(1)   &    0.36(1) \\
                &   0.084697(4)   &    0.37(1)  \\
\hline
 1000-$\infty$  &  0.0065221(1)   &    0.38(1) \\
                &  0.0065222(4)  &     0.39(1) \\
\hline
2000-$\infty$   &  0.00027322(1)  &    0.35(1) \\
                &  0.00027322(1)  &    0.36(1) \\
\hline
\end{tabular}
\end{center}
\caption{Comparison of the LO and $\delta_{q/\bar q\gamma}$
for $pp \to e^+\nu_eX$ in $M_{T, l\nu}$ bins between 
\MCSANC and \cite{Brensing:2007qm}.}
\label{TableDKl}
\end{table}

\begin{table}[!h]
\begin{center}
\begin{tabular}{|c|c|c|}
\hline
$p_{T, l}$/GeV   &   $\sigma_0$/pb  &   $\delta_{q/\bar q\gamma}/\%$\\
\hline
 25-$\infty$    &   4495.8(1)     &    0.059(3) \\
                &   4495.7(2)     &    0.065(1) \\
\hline
 50-$\infty$    &   27.590(1)     &    4.6(1)   \\
                &   27.589(2)     &    4.7(1)   \\
\hline
 100-$\infty$   &   1.7907(1)     &    11.9(1)  \\
                &   1.7906(1)     &    12.3(1)  \\
\hline
 200-$\infty$   &   0.18129(1)   &     16.6(1)  \\
                &   0.18128(1)   &     17.1(1)  \\
\hline
 500-$\infty$   &  0.0065221(1)   &    16.2(1)  \\
                &  0.0065222(4)   &    16.7(1)  \\
\hline
1000-$\infty$   &  0.00027322(1)  &    13.1(1)  \\
                &  0.00027322(1)  &    13.5(1)  \\
\hline
\end{tabular}
\caption{Comparison of the LO cross section and $\delta_{q/\bar q\gamma}$
for $pp \to e^+\nu_eX$ in $p_{T, l}$ bins between 
\MCSANC and \cite{Brensing:2007qm}.}
\label{TableDKll}
\end{center}
\end{table}

In the same setup, in Tables~\ref{Tablellm} and \ref{Tablelm} we present the LO inclusive 
cross sections for processes $pp \to e^-\nu_eX$
and the photon-induced contributions $\delta_{q/\bar q\gamma}$.

\begin{table}[!h]
\begin{center}
\begin{tabular}{|c|c|c|}
\hline
$M_{T, l^-\nu}$/GeV &   $\sigma_0$/pb &   $\delta_{q/\bar q\gamma}/\%$\\
\hline
 50-$\infty$     & 3436.2(1)     & 0.068(2)  \\
\hline
 100-$\infty$    & 20.037(1)     & 0.113(4)   \\
\hline
 200-$\infty$    & 1.08169(1)    & 0.223(2)  \\
\hline
 500-$\infty$    & 0.042127(1)   & 0.328(3)  \\
\hline
 1000-$\infty$   & 0.002584(1)   & 0.349(3)  \\
\hline
 2000-$\infty$   & 0.00008049(1) & 0.344(3)  \\
\hline
\end{tabular}
\caption{The LO results and $\delta_{q/\bar q\gamma}$
for $pp \to e^-\nu_eX$ in $M_{T, l\nu}$ bins.}
\label{Tablellm}
\end{center}
\end{table}

\begin{table}[!h]
\begin{center}
\begin{tabular}{|c|c|c|}
\hline
$p_{T, l^-}$/GeV &   $\sigma_0$/pb &   $\delta_{q/\bar q\gamma}/\%$\\
\hline
 25-$\infty$    & 3436.2(1)     &  0.060(2) \\
\hline
 50-$\infty$    & 20.037(1)     &  5.30(1)  \\
\hline
 100-$\infty$   & 1.0812(1)     & 16.22(2)  \\
\hline
 200-$\infty$   & 0.09503(1)    & 26.31(2)  \\
\hline
 500-$\infty$   & 0.002584(1)   & 34.87(4)  \\
\hline
1000-$\infty$   & 0.00008049(1) & 39.72(2)  \\
\hline
\end{tabular}
\end{center}
\caption{The LO results and $\delta_{q/\bar q\gamma}$
for $pp \to e^-\nu_eX$ in $p_{T, l}$ bins.}
\label{Tablelm}
\end{table}

In Table~\ref{commonTable}
we present the LO cross section for the process $pp \to e^+e^-X$, $\sigma_0$ in pb,
and the corresponding
contributions of photon-induced process $\delta_{q/\bar q\gamma}$ (column 3) and
$\displaystyle\delta_{\gamma\gamma}
 = \displaystyle\frac{\sigma_{\gamma\gamma,0}}{\sigma_0}$ (column 4).
Here we used the setup and input parameters given in ~\cite{Dittmaier:2009cr}.
The results of the \MCSANC integrator are in the first rows,
and the ones from Ref.~\cite{Dittmaier:2009cr} are in the second rows.
Excellent agreement between these two calculations is observed.

In the programs' user interface these corrections are controlled by a new
{\tt iph} flag added to the {\tt iflew} parameter list. Setting {\tt iph=0}
disables photon-induced contributions,
 {\tt iph=1} includes the $q\gamma \to q \ell^+ \ell^-$ (NC DY) and
$q\gamma \to q' \ell^{\pm} \nu$ (CC DY) components, and {\tt iph=2} includes the 
$\gamma\gamma \to \ell^+ \ell^-$ (NC DY) contributions.

\subsection{Higher order corrections}
\label{nTLEC}

In Table~\ref{commonTable} we present the inclusive LO cross section $\sigma_0$ in pb 
for the process $pp \to e^+e^-X$ and the corresponding higher order corrections  
$\displaystyle\delta_{h.o.weak}= \displaystyle\frac{\sigma_{h.o.weak}}{\sigma_0}$
(column 5). The setup and the input parameters are taken from paper~\cite{Dittmaier:2009cr}.
The first rows represent results of the \MCSANC integrator and the second ones
show the numbers computed in~\cite{Dittmaier:2009cr}. 
Again we see excellent agreement between these two calculations.

\begin{table}[!h]
\centering
\begin{tabular}{|l|l|l|l|l|}
\hline
$M_{ll}$/GeV   & $\sigma_0$/pb     &$\delta_{q/\bar q\gamma}/\%$&$\delta_{\gamma\gamma,0}/\%$ &$\delta_{h.o.weak}/\%$\\
\hline
 50-$\infty$  &  738.813(5)       & -0.105(1)  & 0.17(1)  & 0.030(1)     \\
              &  738.773(6)       & -0.11      & 0.17     & 0.030        \\
\hline                                         
100-$\infty$  &  32.7293(2)       &  -0.207(1) & 1.16(1)  &  0.013(1)    \\
              &  32.7268(3)       &  -0.21     & 1.15     &  0.012       \\
\hline                                         
200-$\infty$  &   1.48488(1)      &   0.381(1) & 4.30(1)  &  -0.23(1)    \\
              &   1.48492(1)      &   0.38     & 4.30     &  -0.23       \\
\hline                                         
500-$\infty$  &  0.080942(3)      &  1.522(1)  & 4.92(1)  &  -0.29(1)    \\
              &  0.0809489(6)     &  1.53      & 4.92     &  -0.29       \\
\hline                                                                  
1000-$\infty$ & 0.0067998(1)      & 1.901(1)   & 5.21(1)  &  -0.31(1)    \\
              & 0.00680008(3)     & 1.91       & 5.21     &   -0.31      \\
\hline                                         
2000-$\infty$ & 0.00030375(1)     & 2.343(1)   & 6.18(1)  & -0.31(1)     \\
              & 0.000303767(1)    &  2.34      & 6.17     & -0.32        \\
\hline
\end{tabular}
\caption{
The LO cross section $\sigma_0$ of $pp \to e^+e^-X$ in pb,
the contributions of $(q\gamma$) and ($\gamma\gamma$) configurations in the initial $pp$ state:
$\delta_{q/\bar q\gamma}$ and $\delta_{\gamma\gamma,0}$, 
and higher order corrections $\delta_{h.o.weak}$ ({\tt iho=2}).
Results of \MCSANC (first rows) are compared to
numbers from Ref.\cite{Dittmaier:2009cr} (second rows).}
\label{commonTable}
\end{table}

In the programs' user interface the higher order corrections are controlled by
a new
{\tt iho} flag added to the {\tt iflew} parameter list. Setting {\tt iho=0}
disables higher order correction contributions,
{\tt iho=1} includes linear 
$\left(\Delta \rho - \Delta\rho^{(1)}\Big|^{G_{\mu}}\right)$ term contributions
while {\tt iho=2} includes both linear 
and quadratic $\Delta \rho^2$ term contributions.

\subsection{Forward-backward asymmetry}
\label{nFBA}

The forward-backward asymmetry $A_{FB}^{ff}$ is usually defined as
(see \cite{Buttar:2008jx}):
\bqa
A_{FB}=\frac{F-B}{F+B}
\eqa
where 
\bqa
F=\int_0^1   \frac{d\sigma}{d\cos\vartheta^*}d\cos\vartheta^*,
\qquad
B=\int_{-1}^0 \frac{d\sigma}{d\cos\vartheta^*}d\cos\vartheta^*.
\eqa

The cosine of the angle between the lepton and quark in the $\ell^+ \ell^-$
rest frame is then approximated by 
\bqa
\cos\vartheta^*&=&
 \frac{|p_z(l^+l^-)|}{p_z(l^+l^-)} \cdot
 \frac{2}{m(l^+l^-)\sqrt{m^2(l^+l^-)+p_T^2(l^+l^-)}} \nll
&\times & \left[p^+(l^-)p^-(l^+)-p^-(l^-)p^+(l^+)\right]
\eqa
where $E$ is the energy, $p_z$ and $p_T$ are the longitudinal and transverse components
of the momentum vector, respectively, and $\displaystyle p^\pm=\frac{1}{\sqrt{2}}(E \pm p_z)$.

The \MCSANC results for the forward-backward asymmetries
(Figure~\ref{figpinlo})
were compared with the ones from Ref.~\cite{Buttar:2008jx}. A good agreement was found.

\begin{figure}[!h]
\begin{center}
\begin{tabular}{ll}
\includegraphics[angle=-90,width=0.45\textwidth]{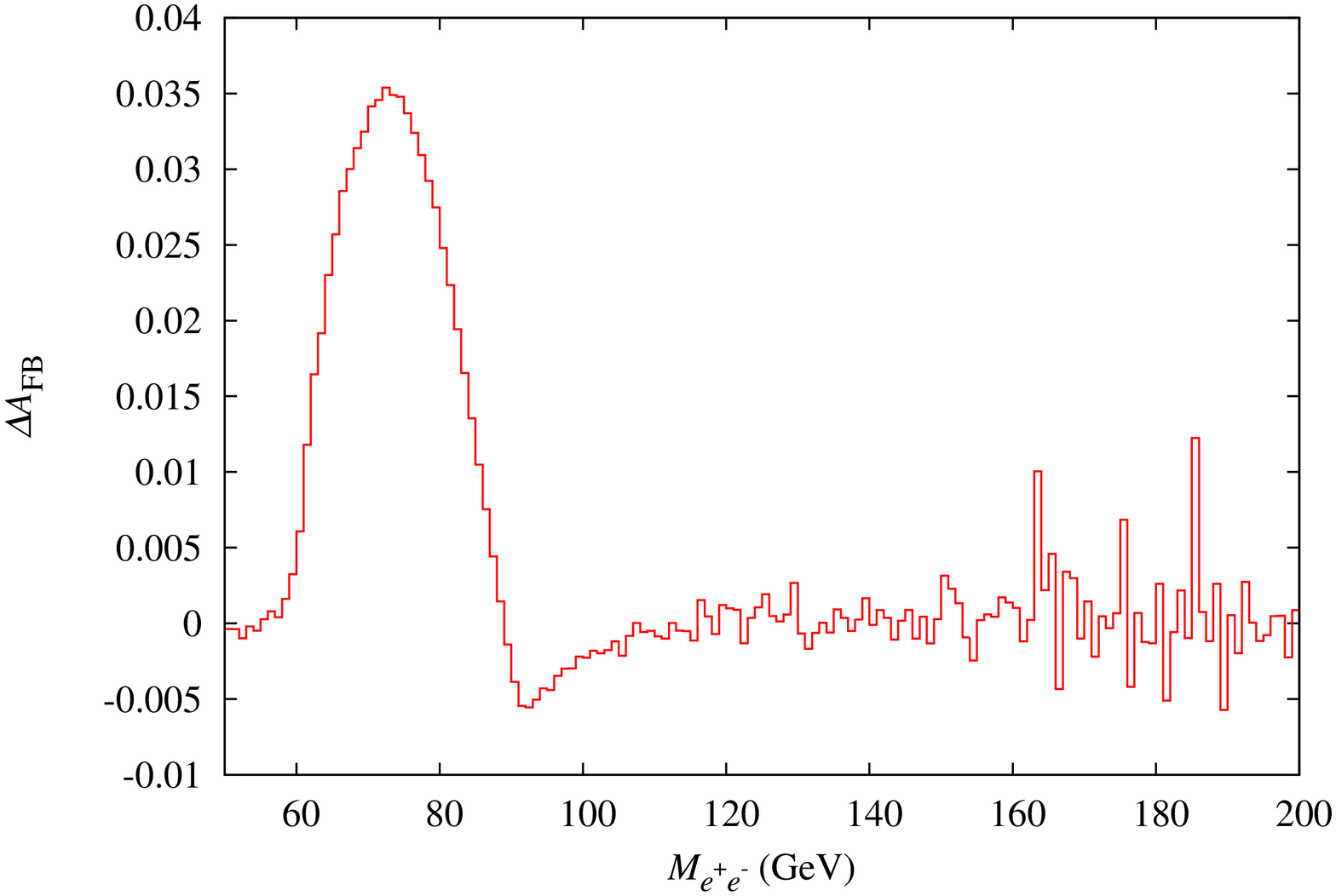} 
~&~
\includegraphics[angle=-90,width=0.45\textwidth]{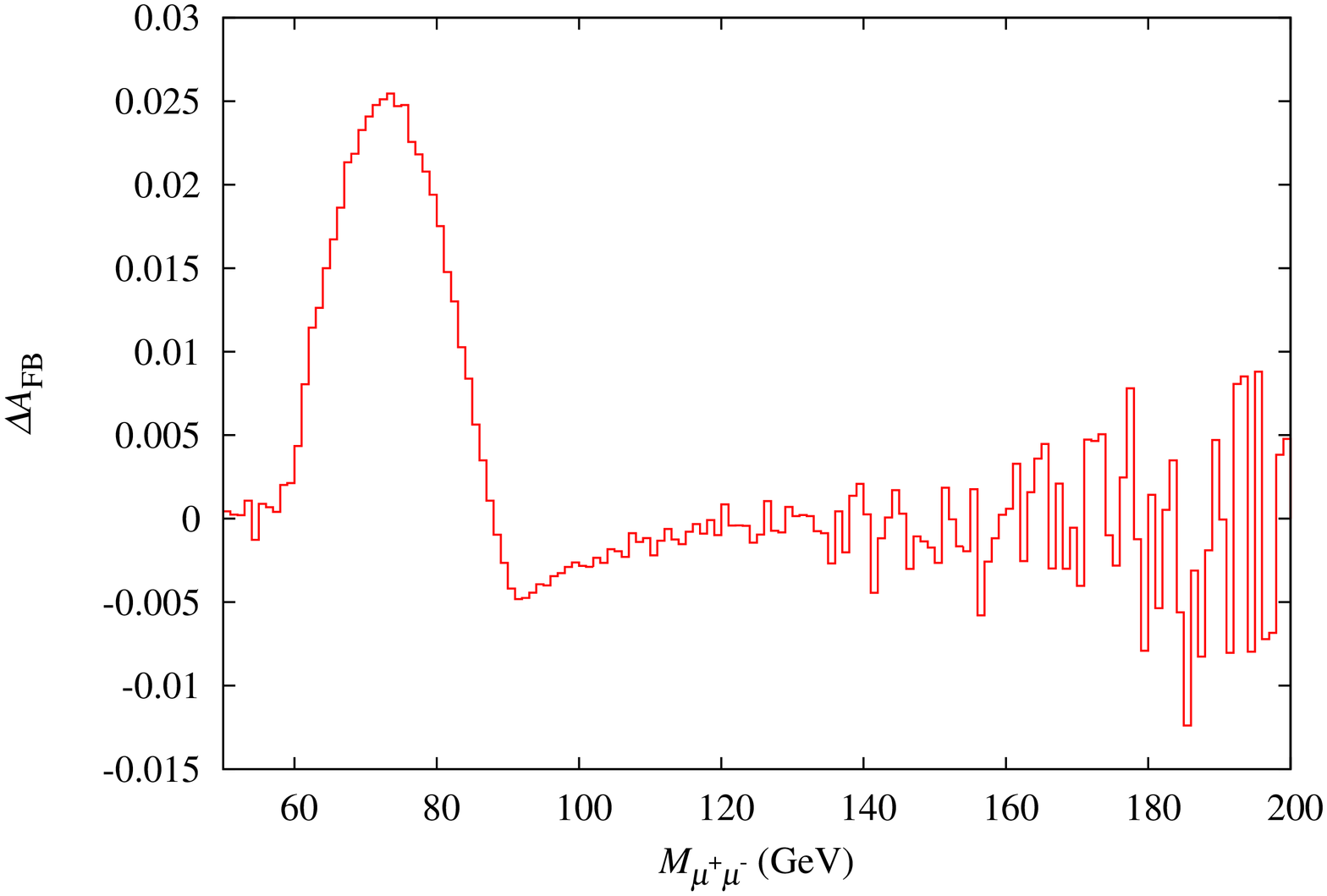}
\end{tabular}
\caption{The forward-backward asymmetry $A_{FB}^{ff}$ of $pp \to \ell^+\ell^-X$ processes
for the $e^+e^-$ (left) and $\mu^+\mu^-$ (right) cases with bare event selection setups.}
\label{figpinlo} 
\end{center}
\end{figure}

\section{Conclusion}
\label{conclusion}
In this paper we have presented an update of the \MCSANC integrator up to {\tt
v.1.20}. The new features include systematic treatment of the photon-induced
contribution in proton--proton collisions and electroweak corrections beyond NLO
approximation. The results of the calculations were compared with the results
of other theoretical groups showing excellent agreement. The current version
of \MCSANC~{\tt v.1.20} is adjusted for studies of various effects due to EW and QCD
radiative corrections to realistic LHC observables.

\section{Acknowledgement}
 We are grateful to W.~von~Schlippe for a critical reading  of this text and useful 
comments.
 The work of L. Rumyantsev was supported by the Institute of Physics theme
 N 213.01-2014/013-VG "Analysis of data and modelling states of the
 near space and the deep space for the purposes of communications and
 navigation".
\bibliographystyle{utphys_spires}
\addcontentsline{toc}{section}{\refname}\bibliography{mcsanc-v120_main}

\end{document}